\documentclass[floatfix,aps,pra,lengthcheck,twocolumn,amsmath,longbibliography]{revtex4-1}

\usepackage{bbm}
\usepackage{esint}
\usepackage{graphicx}
\usepackage[caption=false]{subfig}
\usepackage{ifthen}
\usepackage[colorlinks]{hyperref}
\hypersetup{linkcolor=[rgb]{0.2,0.4,0.7}} 

\newcommand{\pd}{\partial}
\newcommand{\bs}{\boldsymbol}

\newcommand{\diff}[2][]{
\ifthenelse { \equal {#1} {} }
  {\ensuremath{\mathop{\mathrm{d} #2}}}
  {\ensuremath{\mathop{\mathrm{d}^#1 #2}}}
}
\newcommand{\ii}{\mathrm{i}}
\let\oldpi\pi
\renewcommand{\pi}{\mathrm{\oldpi}}
\newcommand{\e}{\mathrm{e}}
\let\derp\perp
\renewcommand{\perp}{{\!\derp\!}}

\begin{document}

\title{Inelastic electron vortex beam scattering}

\author{Ruben Van Boxem}
 \email{ruben.vanboxem@uantwerpen.be}
\author{Bart Partoens}
\author{Jo Verbeeck}
\affiliation{EMAT \& CMT, University of Antwerp,\\Groenenborgerlaan 171, 2020 Antwerpen}

\date{\today}

\begin{abstract}
Recent theoretical and experimental developments in the field of electron vortex beam physics have raised questions on what exactly this novelty in the field of electron microscopy (and other fields, such as particle physics) really provides.
An important part in the answer to those questions lies in scattering theory.
The present investigation explores various aspects of inelastic quantum scattering theory for cylindrically symmetric beams with orbital angular momentum.
The model system of Coulomb scattering on a hydrogen atom provides the setting to address various open questions: How is momentum transferred?
Do vortex beams selectively excite atoms, and how can one employ vortex beams to detect magnetic transitions?
The analytical approach presented here, provides answers to these questions.
OAM transfer is possible, but not through selective excitation; rather, by pre- and post-selection one can filter out the relevant contributions to a specific signal.
\end{abstract}

\maketitle

\section{Introduction}

The field of phase vortices in beams has a relatively short but vibrant history, beginning in optics and ending in all forms of wave mechanical systems.
A good overview of theory and applications in optics can be found in refs.~\cite{Allen,Dennis}.
Interesting aspects include the quantization of orbital angular momentum (OAM)~\cite{Andersen}, the opto-mechanical effects of these beams~\cite{Andersen,Simpson}, quantum information, computing, and communication uses~\cite{Boyd,Eilenberger}, and of course scattering applications~\cite{Matula,Ivanov,Ivanov_phase}.
Various ways of obtaining electron vortex beams have been demonstrated~\cite{Schattschneider_fork,Schattschneider_mode_conversion,Verbeeck,McMorran,Clark_Lauraperture,Beche,Grillo,Krivanek} since their inception on paper in 2007~\cite{Bliokh_semiclassical} and experimentally in 2010~\cite{Uchida,Verbeeck}.
Several investigations into the scattering aspects of electron vortex beams have been published~\cite{Rusz,Schattschneider_mapping}, and this work aims to add to that list a thorough and general analysis of inelastic electron vortex scattering.

The main questions we investigate in this paper all pertain to the detection of and mechanism of OAM transfer of a fast vortex electron to an atomic system.
The results presented in this paper describe the result of a detailed investigation into the quantum dynamics of OAM transfer from an incoming electron vortex beam to an atomic electron.
It is an attempt to describe what the incoming and outgoing OAM and transverse momentum do to the scattering amplitude, and how these can be exploited in discovering what happened during a scattering event.
In what follows, inelastic scattering of fast electron Bessel beams is calculated for the model system of a hydrogen atom, of which several transition amplitudes are explicitly calculated.
Although we consider a spherically symmetric atomic system, we prefer to discuss everything relevant to the beam axis, and we treat transitions with OAM transfer in that view.
Special detail will be given to how and when selection rules are present, and if they are not, we describe the physical reasons behind their vanishing outside of special circumstances.
We put the emphasis on what vortex beams contribute to the scattering, and what they do not.
To this effect, two alternative approaches are pursued, allowing for a sufficiently nuanced and accurate report on the properties of electron vortex beam atomic excitations.
Critical are the concepts of pre- and post-selection on the scattered wave, and special attention is given to transitions in which the beam's OAM is transfered to the atomic electron.
Effects related to relativity, spin or recoil of the nucleus are ignored.
This treatment is also more simplistic than the more advanced treatments based on the density matrix~\cite{Schattschneider_density_matrix,Schattschneider_mdff}, and relate more closely to the earlier theoretical advances in inelastic electron-atom scattering~\cite{Inokuti,Egerton_lorentzian,Egerton_oscillator_strength}.

The next section will quickly review various inelastic scattering concepts, and puts special emphasis on the details often omitted in a textbook treatment of quantum scattering on atomic systems which are important here.
The third section presents the two methods developed to attack this problem analytically.
The fourth section is a discussion of the most important results.
The final section presents our conclusions.

\subsection*{Notation}

In order to remove clutter from crucial steps in the calculation, strict notational conventions are defined here.
Real space coordinates are denoted $\bs r = (x,y,z) = (r_\perp,\varphi,z)$.
Basis vectors are written as $\bs e_i$ where $i$ is the relevant coordinate.
Momentum space coordinates are denoted $\bs k = (k_x, k_y, k_z) = (k_\perp, \phi, k_z)$.
It is often advantageous to split any 3D vector $\bs v$ into a $z$-component $v_z$ and a perpendicular component $\bs v_\perp = (v_x, v_y)$ with size $v_\perp$.
Primed variables are denoted $\bs r^\prime = (x^\prime, y^\prime, z^\prime) = (r_\perp^\prime, \varphi^\prime, z^\prime)$ and similar for $\bs k^\prime$.
Partial derivatives are denoted $\pd_x = \frac{\pd}{\pd x}$ and are taken to act on everything to the right of the symbol.
Primed coordinates and quantum numbers refer to the final (outgoing) state.
Momentum transfer is denoted $\bs q$ and is equal to $\bs k - \bs k^\prime$.
OAM transfer is denoted $\Delta m$ and is equal to $m - m^\prime$.


\section{A cylindrical view on inelastic electron plane wave scattering}

\subsection{Inelastic scattering of an electron on an atom}

In this section, various entities are defined and plane wave scattering theory is quickly reviewed.
Certain often forgotten aspects which are important to the cylindrical wave case are brought forward.

\subsubsection{Interaction potential}

The interaction potential for an electron scattering on a single-electron atomic system is given by
\begin{equation} \label{eq:inelastic_coulomb_potential}
 V = e^2 \left( \frac{1}{|\bs r-\bs r^\prime|} - \frac{Z}{r^\prime} \right).
\end{equation}
Here, $\bs r^\prime$ is the lab coordinate for the scattering electron with respect to the nucleus, and $\bs r$ is the atomic electron's coordinate relative to the nucleus (which is located at the origin).
Extending this to multiple single-electron atomic states is formally trivial: summing over every electron's coordinate suffices, so that $\bs r \rightarrow \sum \bs r_j$.
The relative distance $|\bs r - \bs r^\prime|$ will complicate the cylindrical scattering treatment in the next sections.

\subsubsection{Single electron atomic wave functions}

To make the role of the atomic system's OAM apparent, we assume that the (projected) OAM is a good quantum number, and separate the atomic wave functions into an azimuthal part and the remaining $(r_\perp,z)$ part:
\begin{align} \label{eq:atomic_azimuthal}
 &|i\rangle = \frac{\e^{\ii m \varphi}}{\sqrt{2\pi}} |\alpha\rangle, &|f\rangle = \frac{\e^{\ii m^\prime \varphi}}{\sqrt{2\pi}} |\beta\rangle.
\end{align}
For a hydrogen-like atom with (effective) nuclear charge $Z$ and atomic radius $a_\mu = a_0 m_e / \mu$ ($\mu$ is the reduced mass of the atomic system), the orthonormal wave functions have the following form:
\begin{equation} \label{eq:hydrogen_wave_functions}
 \langle \bs r | nlm \rangle = \sqrt{\eta^3 \frac{(n-l-1)!}{2n (n+1)!}}\frac{\e^{\ii m \varphi}}{\sqrt{2\pi}}  (\eta r)^l L_{n-l-1}^{2l+1}(\eta r) \e^{-\frac{\eta r}{2}},
\end{equation}
Where $\eta = 2 Z / (n a_\mu)$, and $n = 1, 2, 3 \hdots$, $l = 0, \hdots n-1$, and $m = -l, \hdots l$ are the radial, angular and magnetic quantum numbers, respectively.
The OAM of a state, determined by the operator $\hat{L}$, is given by $l$, and the projected OAM, given by $\hat{L_z}$ and the one of primary interest here, is given by $m$.
In what follows, we shall assume $\mu\approx m_e$ and so $a_\mu \approx a_0$, but keep $Z$ so that any dependence on atomic number is clear.

\subsubsection{Scattering amplitude}

The inelastic scattering amplitude in the first Born approximation can be written as follows:
\begin{equation} \label{eq:inelastic_scattering_amplitude}
 f_{fi}[\bs k^\prime, \Phi] = -\frac{2m_e N}{4\pi \hbar^2} \langle \bs k^\prime | \langle f | V | i \rangle | \Phi \rangle,
\end{equation}
where $N$ takes care of the normalization of the scattering electron's states~\cite{VanBoxem_Rutherford}, $|i\rangle$ and $|f\rangle$ are the initial and final atomic states, $|\Phi\rangle$ is the incoming state (traditionally taken to be a plane wave).
The outgoing plane wave is defined as:
\begin{equation} \label{eq:plane_wave}
 \langle \bs r | \bs k^\prime \rangle = \frac{\e^{\ii \bs k^\prime \cdot \bs r}}{(2\pi)^{3/2}}.
\end{equation}
The scattering amplitude determines the modulation of the outgoing (scattered) spherical wave so that the total wave function obeys the following relation:
\begin{equation}
 \Psi(\bs r) = \Phi(\bs r) + f_{fi}[\bs k^\prime, \Phi] \frac{\e^{\ii \bs k \cdot \bs r}}{r}.
\end{equation}
The scattering amplitude, Eq.~\eqref{eq:inelastic_scattering_amplitude}, depends on the atomic states involved in a certain transition, and thus the probability of scattering an electron in a certain direction depends on these states.

Filling in Eq.~\eqref{eq:inelastic_scattering_amplitude} with Eq.~\eqref{eq:inelastic_coulomb_potential}, we obtain:
\begin{align}
 f_{fi}^\mathrm{B} [\bs k^\prime, \Phi] &= -\frac{m_e e^2 N}{2\pi\hbar^2} \left( \langle \bs k^\prime | \langle f | \frac{1}{|\bs r - \bs r^\prime|} | i \rangle | \Phi \rangle \right. \notag \\ 
 &\phantom{-\frac{m_e e^2 N}{2\pi\hbar^2} = \Bigl(} \left.- Z \langle f | i \rangle \langle \bs k^\prime | \frac{1}{r^\prime} | \Phi \rangle \right).
 \label{eq:inelastic_scattering_amplitude_full}
\end{align}

\subsubsection{The final state} \label{sec:final_state}

\begin{figure}
 \centering
 \includegraphics[width=\linewidth]{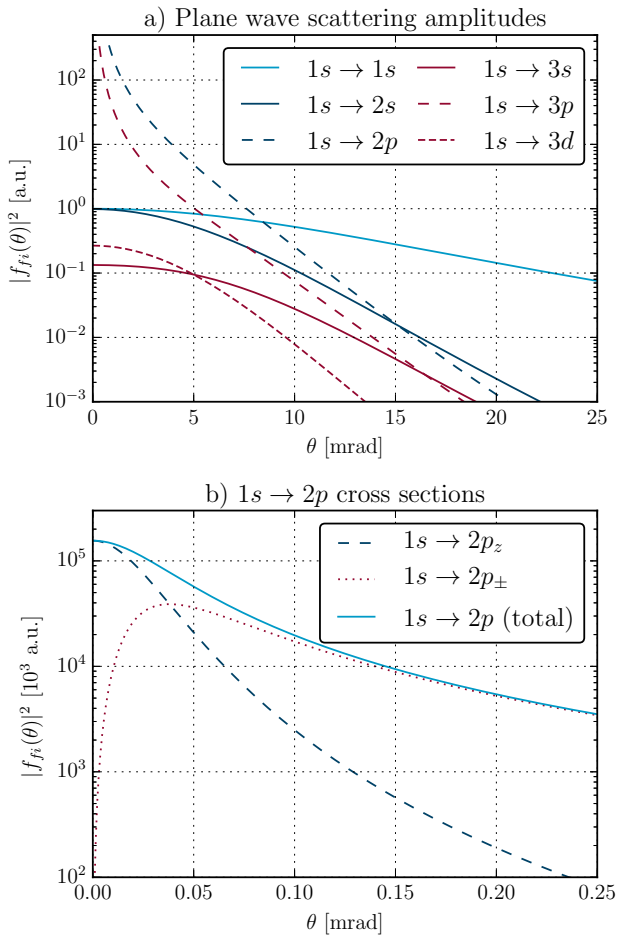}
 \caption{(color online) a) Plane wave scattering amplitudes for 120 keV electrons for various transitions from the ground state. One can see the relative size of the amplitudes shifts depending on the exact scattering angle in question. b) Separate cross sections for $2p_z$ and $2p_\pm$ excitation. The typical Lorentz profile appears~\cite{Schattschneider_magic_angle}. The total $1s\rightarrow 2p$ cross section is the sum of the other three. Note the lower $\theta$ scale is a factor 100 times smaller than the one in the top graph. This tiny scale is due to the relatively small $\Delta E$ for pure hydrogen. More realistic systems will show the same features at a much larger scale. \label{fig:final_state}}
\end{figure}
The outgoing electron's momentum, $k^\prime$ is related to the incoming electron's momentum $k$ through the energy transferred to the atomic system, $\Delta E$, as follows:
\begin{equation} \label{eq:outgoing_momentum}
 k^{\prime 2} = k^2 - \frac{2 m_e}{\hbar^2} \Delta E.
\end{equation}
The scattering angle $\theta$ can be introduced by substituting:
\begin{equation}
 q^2 = k^2 + k^{\prime 2} -2 k k^\prime \cos{\theta}.
\end{equation}
Using the above with Eq.~\eqref{eq:inelastic_scattering_amplitude_full} and Eq.~\eqref{eq:hydrogen_wave_functions}, one can plot the angular dependence of the scattering amplitudes for various hydrogen transitions.
These are shown in Fig.~\ref{fig:final_state}a.
The different final states (s/p/d/\ldots) result in specific angular regions in which some dominate or are suppressed with respect to others, which implies some transitions can be filtered out roughly by post-selection on $\theta$ (which directly corresponds to outgoing transverse momentum).
This can be exploited to map anisotropic bonding in crystals~\cite{Jouffrey,Schattschneider_magic_angle}.

\begin{figure}
 \centering
 \includegraphics{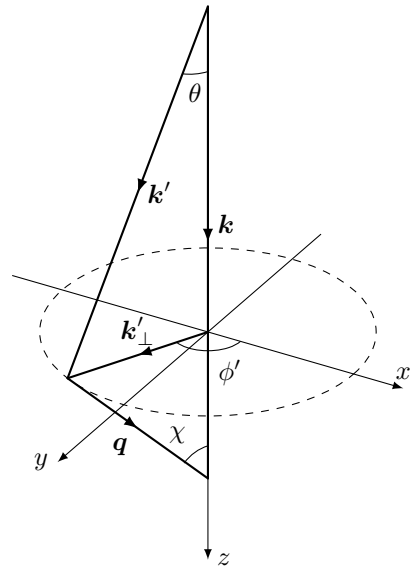}
 \caption{For plane wave scattering, the orientation of the final state depends on the scattering angle $\theta$ and the transition energy $\Delta E$ as shown in Eq.~\eqref{eq:final_state_tilt_angle}. The figure also shows the azimuthal angle, $\phi^\prime$ of the outgoing wave with respect to a fixed set of axes. \label{fig:momentum_transfer_orientation}}
\end{figure}
We will now consider a ground state excitation to a state with fixed $(n^\prime, l^\prime)$, \textit{e.g.} the $2p$ orbitals, and see if the outgoing wave's OAM can be used to provide more details of the final state.
The question we ask ourselves is: what determines scattering to a particular (projected) OAM?
This example final state consists of three substates: $2p_z$, $2p_+$, and $2p_-$.
The orientation of these final states is determined by the direction of the  $z^\prime$ axis, along which their angular momentum is projected to give quantum number $m^\prime$.
This is crucial to the whole analysis.
Using the Fourier translation theorem, the plane wave scattering amplitude becomes
\begin{equation}
f_{fi}^\mathrm{PW}(\bs q) = - \frac{2 m_e e^2}{\hbar^2} \langle f | \frac{\e^{\ii \bs q \cdot r^\prime} - Z}{q^2} | i \rangle.
\end{equation}
This can be calculated in position space by aligning the $z^\prime$ axis along the momentum transfer vector $\bs q$:
\begin{equation} \label{eq:plane_wave_scattering_amplitude}
f_{fi}^\mathrm{PW}(\bs q) = - \frac{2 m_e e^2}{\hbar^2} \int \diff[3]{\bs r^\prime} \psi_f^*(\bs r^\prime) \frac{\e^{\ii q r^\prime \cos{\theta^\prime}} - Z}{q^2} \psi_i(\bs r^\prime).
\end{equation}
The final states are thus automatically quantized along the direction of $\bs q$, when expressed in the $\bs r^\prime$ coordinate system.
\footnote{
 By expanding the exponential in spherical functions, separates various contributions to the scattering amplitude into many terms. These were calculated in e.g. Ref.~\cite{Loffler_transition}.
}
Furthermore, due to the integration over the azimuthal coordinate $\varphi^\prime$ in Eq.~\eqref{eq:plane_wave_scattering_amplitude}, only the tilted states for which $\Delta m = 0$ contribute.
In our example of ground state excitation, this means that for a certain $(n^\prime,l^\prime)$, only the state with $m^\prime=0$ gives a non-zero contribution.
Indeed, the transition amplitude for $1s\rightarrow 2p$ is given by following expression:
\begin{equation} \label{eq:plane_wave_scattering_amplitude_1s2p}
_{\chi,\phi^\prime}\langle 2p_z | \frac{\e^{\ii \bs q \cdot \bs r}}{q^2} | 1s \rangle = \left(\frac{Z}{a_0}\right)^5\frac{12\ii \sqrt{2}}{q \left[q^2 +  \left(\frac{3Z}{2a_0}\right)^2 \right]^3}.
\end{equation}
The subscript $\chi$ expresses that the quantization axis is tilted over an angle of $\chi$, defined by
\begin{equation} \label{eq:final_state_tilt_angle}
 \tan{\chi} = \frac{q_\perp}{q_z} = \frac{k^\prime \sin{\theta}}{k - k^\prime \cos{\theta}}, 
\end{equation}
and rotated around the beam's $z$ axis by an angle $\phi^\prime$.
This scattering geometry is shown in Fig.~\ref{fig:momentum_transfer_orientation}.
The reader can verify that Eq.~\eqref{eq:plane_wave_scattering_amplitude_1s2p} is zero for the other tilted substates (in our example, these are $2p_+$ and $2p_-$).
This tilted state ($z^\prime$ is rotated with respect to the beam's $z$ axis) can be projected onto untilted states which are quantized along the beam direction.

As the atomic orbitals are determined by a radial function multiplied with an orbital angular momentum eigenstate $|l,m\rangle$, their rotation properties are fully determined by the Wigner D-matrix:
\begin{equation}
 \langle l, m | \hat{R}(\alpha, \beta, \gamma) | l^\prime, m^\prime \rangle = \delta_{l^\prime, l} D_{m, m^\prime}^{(l)}(\alpha, \beta, \gamma).
\end{equation}
Rotating an angular momentum eigenstate over the angles $\alpha$, $\beta$, and $\gamma$ (in the z-y-z convention) turns it into a sum of states of equal $l$, but all $m$:
\begin{equation} \label{eq:angular_momentum_rotation}
 \hat{R}(\alpha,\beta,\gamma) |l, m^\prime \rangle = \sum_{m=-l}^{+l} D_{m, m^\prime}^{(l)}(\alpha, \beta, \gamma) |l, m \rangle .
\end{equation}
For the rotation of an $m^\prime=0$ state oriented along the vector $\bs q$ as in Fig.~\ref{fig:momentum_transfer_orientation}, the matrix elements take on the following form:
\begin{equation} \label{eq:wigner_d_matrix_0}
 \begin{aligned} 
  D_{m,0}^l(\phi^\prime, \chi, 0) &= \e^{-\ii m \phi^\prime} d_{m,0}^l(\chi) \\
  &= \e^{-\ii m \phi^\prime} \sqrt{\frac{(l-m)!}{(l+m)!}} P_l^m (\cos{\chi}),
 \end{aligned}
\end{equation}
where $P_l^m$ are the associated Legendre polynomials.

As an example, consider the $1s\rightarrow2p$ transitions of the hydrogen atom, which can readily be calculated by filling in Eq.~\eqref{eq:plane_wave_scattering_amplitude} with Eq.~\eqref{eq:hydrogen_wave_functions}.
With respect to the beam direction, the excited $2p_z$ state is tilted by the angle $\chi$ as shown in fig.~\ref{fig:momentum_transfer_orientation}.
Applying Eqs.~\eqref{eq:angular_momentum_rotation} and~\eqref{eq:wigner_d_matrix_0}, for this example, one obtains:
\begin{subequations} \label{eq:p_state_rotation}
 \begin{equation}
  | 2p_z \rangle_{\chi,\phi^\prime} = \frac{\sin{\chi}}{\sqrt{2}} \left( \e^{\ii \phi^\prime} | 2p_- \rangle - \e^{-\ii \phi^\prime} | 2p_+ \rangle \right) + \cos{\chi} | 2p_z \rangle.
 \end{equation}
 Here,
 \begin{align} \label{eq:chi_in_function_of_q}
  &\sin{\chi} = \frac{q_\perp}{q}, &\cos{\chi} = \frac{q_z}{q},
 \end{align}
\end{subequations}
This expression gives the projection of the tilted state $|2p_z\rangle_{\chi,\phi^\prime}$ into its untilted components (quantized with respect to the beam's $z$ direction).

The explicit form of the projection of the $\bs q$-oriented $2p_z$ state onto a beam axis-oriented $2p$ state, \textit{e.g.} $\langle 2p_\pm | 2p_z \rangle_{\chi, \phi^\prime}$, in Eq.~\eqref{eq:p_state_rotation}, allows us to write the scattering amplitude to \textit{e.g.} the $2p_\pm$ states directly using Eq.~\eqref{eq:plane_wave_scattering_amplitude_1s2p}:
\begin{align} \label{eq:plane_wave_1s2pplusmin}
 f_{1s\rightarrow 2p_\pm}^\mathrm{PW} &= -\frac{2m_e e^2}{\hbar^2} \langle 2p_\pm | \frac{\e^{\ii \bs q \cdot \bs r}}{q^2} | 1s \rangle \notag \\
 &=-\frac{2m_e e^2}{\hbar^2} \langle 2p_\pm | 2p_z \rangle_{\chi,\phi^\prime} \!\times\! {}_{\chi,\phi^\prime} \langle 2p_z | \frac{\e^{\ii \bs q \cdot \bs r}}{q^2} | 1s \rangle \notag \\
 &= \left(\frac{Z}{a_0}\right)^5 \e^{\mp \ii \phi^\prime} \frac{-12\ii m_e e^2 q_\perp}{\hbar^2 q^2  \left[q^2 + \left( \frac{3Z}{2a_0}\right)^2\right]^3}.
\end{align}
This is obtained by inserting a complete $l=1$ basis and remarking that the transition element is only non-zero for $|2p_z\rangle_{\chi,\phi^\prime}$, as discussed above.
Note that the outgoing wave has acquired the opposite OAM of the excited atomic state, and thus OAM has been transferred to the bound electron.
The above result is not new, but it usually appears as a cross section.
More specific forms are exploited in \textit{e.g.} momentum resolved EELS experiments~\cite{Calmels,Klie}, where dynamical diffraction also plays a large role in the final distribution of scattered electrons.
The $2p_z$ excitation, where no OAM is transferred, can be calculated directly as well:
\begin{equation} \label{eq:plane_wave_1s2pz}
 f_{1s\rightarrow 2p_z}^\mathrm{PW} = \left(\frac{Z}{a_0}\right)^5 \frac{-12\sqrt{2} \ii m_e e^2 q_z}{\hbar^2 q^2 \left[q^2 + \left(\frac{3Z}{2a_0}\right)^2 \right]^3}.
\end{equation}
Both Eq.~\eqref{eq:plane_wave_1s2pplusmin} and~\eqref{eq:plane_wave_1s2pz} are shown in Fig.~\ref{fig:final_state}b.
These scattering amplitudes can be separately observed if one filters the outgoing wave on its OAM.
Techniques to achieve this for electromagnetic waves exist~\cite{Berkhout_Sorting,Saitoh}, but in electron optics these are still in development~\cite{Guzzinati,Clark_Quantitative}.

\subsection{The vortex beam basis state: Bessel beams}

The simplest form of vortex beams is provided by the solution of the Schr\"odinger equation in cylindrical coordinates:
\begin{equation} \label{eq:bessel_beam}
\langle \bs r | \bs k, \ell \rangle = \psi_{\bs k, \ell}(\bs r) = \frac{\e^{\ii \ell \varphi}}{\sqrt{2\pi}} J_\ell(k_\perp r_\perp) \frac{\e^{\ii k_z z}}{\sqrt{2\pi}}.
\end{equation}
This exact solution encompasses the beam features that interest us: the quantized (projected) OAM $\hbar \ell$, and the longitudinal and transverse momenta $\hbar k_z$ and $\hbar k_\perp$.
One can define an \emph{opening angle} as $\alpha = \tan^{-1}(k_\perp/k_z)$ which is the angle the momentum vector makes with the $z$ axis.
The energy of this field-free state is independent of its OAM: $E = \hbar^2 (k_z^2 + k_\perp^2) / (2m)$.
The Bessel beam can be written in terms of its momentum components, which shows that this state is a ring of tilted plane waves in momentum space:
\begin{equation} \label{eq:bessel_fourier_representation}
\psi_{\bs k, \ell}(\bs r) = (-\ii)^\ell \int \frac{\diff{\phi}}{(2\pi)^2} \e^{\ii \ell \phi} \e^{\ii \bs k \cdot \bs r}.
\end{equation}
This representation was used to calculate the elastic Coulomb scattering amplitude in an earlier work~\cite{VanBoxem_Rutherford}, of which the results will be useful here in simplifying some of the equations.

Bessel beams are basis states, much like plane waves.
It is impossible to create true Bessel beams, even in a lab setting.
Only approximations can be realized, which admittedly show the various properties of real Bessel beams in a limited way~\cite{Bouchal}.
Nonetheless, they provide a good basis to calculate scattering amplitudes because they encode features such as cylindrical symmetry, convergence angle (transverse momentum), and OAM in a natural way.
Note that for a convergent beam, the wave function of the probe in real space is nothing more than a coherent superposition of Bessel beam basis states, integrated over the aperture radius.
For a non-vortex probe, this reads:
\begin{equation}
\Psi(\bs r) \propto \int_0^\infty \diff{k_\perp} A(k_\perp) k_\perp J_0(k_\perp r_\perp).
\end{equation}
Here, $A(k_\perp)$ describes the aperture (for example as a step function).
This is nothing new, and often called the Hankel transform.
What this means is that the scattering amplitude of a real convergent beam can be obtained from that of a Bessel beam by integrating over a $k_\perp$ range given by the aperture radius.
For a vortex probe, one can write:
\begin{equation}
\Psi_\ell (\bs r) \propto \e^{\ii \ell \varphi} \int_0^\infty \diff{k_\perp} A(k_\perp) k_\perp J_\ell(k_\perp r_\perp),
\end{equation}
which is nothing more than the $\ell$-th order Hankel transform~\footnote{See e.g. p. 706 in ref.\cite{Bronshtein}}.

\section{Inelastic vortex beam scattering} \label{sec:relative_coordinate_problem}

\subsection{General formulation} \label{sec:general_formulation}

We consider a perfectly centered vortex, so that the atom lies exactly on the beam's OAM axis.
Beam displacement can be taken into account by using the Bessel addition formula (see Eq.~\eqref{eq:bessel_addition_theorem}) for the incoming beam in the transverse plane.
Any non-zero displacement of the beam (which is to be expected in any realistic situation) will thus introduce a progressively larger number of OAM modes significantly contributing to the scattering.
The extra modes contribute with a factor determined by the transverse momentum, displacement, and the OAM of that contribution through $J_{\ell-\mu}(k_\perp r_{0 \perp})$, where $\ell$ is the pure OAM mode displaced over a distance $r_{0 \perp}$, and $\mu$ is the OAM of a mode introduced by the displacement.
For $r_{0\perp} = 0$, all other contributions disappear.
These additional OAM modes will coherently contribute to the final scattering amplitude, and thus interfere upon calculating the differential cross section.

Following up on ref.~\cite{VanBoxem_Rutherford}, one can calculate inelastic scattering amplitudes by replacing $|\Phi\rangle$ in Eq.~\eqref{eq:inelastic_scattering_amplitude_full} with a Bessel beam $|\bs k, \ell\rangle$, Eq.~\eqref{eq:bessel_beam}.
Ignoring the (purely elastic) term (which was calculated in ref.~\cite{VanBoxem_Rutherford} and repeated in App.~\ref{app:elastic_scattering_amplitude}), this leads to:
\begin{equation} \label{eq:inelastic_bessel_beam_scattering_amplitude}
 f_{fi}^\mathrm{V} = -\frac{m_e e^2}{2\pi \hbar^2}
 \begin{aligned}[t]
  &\int \diff[3]{\bs r} \psi_f^*(\bs r) \psi_i(\bs r) \\
  &\times \int \diff[3]{\bs r^\prime} \e^{-\ii \bs k^\prime \cdot \bs r^\prime} \frac{1}{|\bs r - \bs r^\prime|} \e^{\ii \ell \varphi^\prime} J_\ell(k^{}_\perp r_\perp^\prime) \e^{\ii k_z z}.
 \end{aligned}
\end{equation}
The relative atomic coordinate of the atomic electron, $\bs r^R = \bs r - \bs r^\prime$, plays a central role in how to solve this problem.
We would like to substitute this in the inner integral as to not involve the atomic wave functions already.
The relative distance $|\bs r - \bs r^\prime|$ is difficult to substitute directly in this expression, due to the presence of the transverse coordinate.
There are two ways to proceed: use the Bessel addition theorem to mathematically displace the Bessel beam to this relative coordinate, or introduce the Fourier representation of the Bessel beam, Eq.~\eqref{eq:bessel_fourier_representation}.
Both methods lead to interesting physical insights, and thus both are treated below in detail.

Note that much like the elastic results presented in a previous article~\cite{VanBoxem_Rutherford}, any of these scattering amplitudes can be summed/integrated over a certain range of $k_\perp$, $k_z$ and/or $\theta$ to more closely resemble real wave packets.
This allows for modelling of a focused electron probe with a specific convergence angle (including annular apertures), or a certain collector angle range for comparison with the experiment.

\subsection{Displaced Bessel beam representation} \label{sec:bessel_displacement}

\begin{figure}
 \centering
 \includegraphics{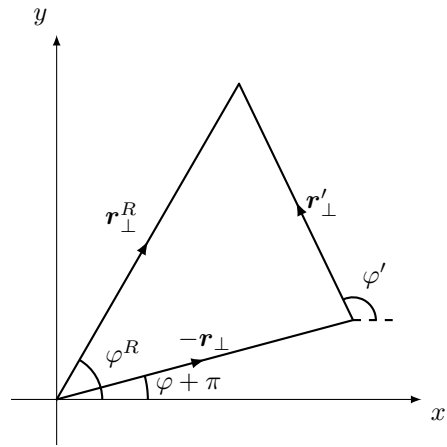}
 \caption{In-plane geometry as used in Eq.~\eqref{eq:bessel_addition_theorem} for the shift to the atomic electron's coordinate. \label{fig:bessel_displacement}}
\end{figure}
\begin{figure}
 \centering
 \includegraphics{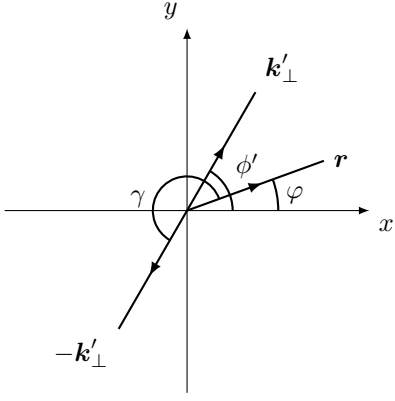}
 \caption{Geometry for the outgoing plane wave's Jacobi-Anger expansion in Eq.~\eqref{eq:inelastic_jacobi_anger}. Here, $\gamma$ is the angle between $-\bs k_\perp$ and $\bs r_\perp$. \label{fig:inelastic_jacobi_anger}}
\end{figure}

The relative coordinate problem described above can be worked around by applying the Bessel addition theorem, with which a displaced Bessel beam can be replaced with a series of weighted centred Bessel functions over all orders.
In our case, the theorem states~\footnote{\href{http://dlmf.nist.gov/10.23\#E7}{Eq. 10.23.7} in ref.\cite{DLMF}}:
\begin{equation} \label{eq:bessel_addition_theorem}
 \begin{aligned}
  &\frac{\e^{\ii k_z z^\prime}}{\sqrt{2\pi}} J_\ell(k^{}_\perp r_\perp^\prime) \frac{\e^{\ii \ell \varphi^\prime}}{\sqrt{2\pi}} \\
  &=\e^{\ii k_z z} \sum_{\mu=-\infty}^{+\infty} J_\mu(k_\perp r_\perp) \e^{\ii \mu \varphi} \frac{\e^{\ii k_z z^R}}{\sqrt{2\pi}} J_{\ell-\mu}(k^{}_\perp r_\perp^R) \frac{\e^{\ii (\ell - \mu) \varphi^R}}{\sqrt{2\pi}},
 \end{aligned}
\end{equation}
where the relative coordinate is denoted $\bs r^R = \bs r - \bs r^\prime$.
This is illustrated in Fig.~\ref{fig:bessel_displacement}.

Applying this to Eq.~\eqref{eq:inelastic_bessel_beam_scattering_amplitude}, one obtains:
\begin{align}
 f_{fi}^\mathrm{V} (\bs k^\prime; \bs k, \ell) &= -\frac{m_e e^2 N}{2\pi \hbar^2} \sum_{\mu=-\infty}^{+\infty} 
 \begin{aligned}[t]
  &\langle f | \e^{-\ii \bs k^\prime \cdot \bs r} J_\mu(k_\perp r_\perp) \e^{\ii \mu \varphi} | i \rangle \\
  &\times \langle \bs k^\prime | \frac{1}{r^R} | \bs k, \ell - \mu \rangle
 \end{aligned} \notag \\
 &= -\sum_{\mu=-\infty}^{+\infty} 
 \begin{aligned}[t]
  &f_\mathrm{el}^V(\bs k^\prime; \bs k, \ell - \mu) \\
  &\times \langle f | \e^{-\ii \bs k_\perp^\prime \cdot \bs r_\perp} J_\mu(k_\perp r_\perp) \e^{\ii \mu \varphi} \e^{\ii q_z z} | i \rangle
 \end{aligned} \notag \\
 &= -\sum_{\mu=-\infty}^{+\infty} f_\mathrm{el}^V(\bs k^\prime; \bs k, \ell - \mu) \widetilde{\mathcal{M}}_{fi}(\bs k^\prime, \bs k, \mu) \label{eq:inelastic_scattering_amplitude_cylindrical}
\end{align}
In the above expressions, $f_\mathrm{el}^V$ is the Coulomb scattering amplitude as given in App.~\ref{app:elastic_scattering_amplitude}.
This form resembles the plane wave result quite closely: a Coulomb scattering factor ($1/q^2$ for plane wave incidence, assuming the form of $f_\mathrm{el}^V$ here) multiplied by a matrix element $\widetilde{\mathcal{M}}_{fi}$ over the two bound states involved.
This matrix element will take on a simpler form below, although the functional form of the ``matrix element operator" is not the same as in the plane wave case, \textit{i.e.} it is not simply $\e^{\ii \bs q \cdot \bs r}$.

The outgoing plane wave can be written as a series of Bessel functions using the Jacobi-Anger theorem (see also Fig.~\ref{fig:inelastic_jacobi_anger}):
\begin{equation} \label{eq:inelastic_jacobi_anger}
 \e^{-\ii \bs k_\perp^\prime \cdot r^{}_\perp} = \sum_{\lambda=-\infty}^{+\infty} (-\ii)^\lambda J_\lambda(k_\perp^\prime r^{}_\perp) \e^{\ii \lambda (\phi^\prime - \varphi)},
\end{equation}
which makes the remaining hidden azimuthal angular dependence explicit, exposing the possibility of selection rules on (projected) OAM quantum numbers.
The matrix element then becomes:
\begin{equation} \label{eq:inelastic_matrix_full}
 \widetilde{\mathcal{M}}_{fi} = \sum_{\lambda=-\infty}^{+\infty}
 \begin{aligned}[t]
  &\left(-\ii \e^{\ii \phi^\prime} \right)^\lambda  \\
  &\times \langle f | J_\mu(k^{}_\perp r^{}_\perp) J_\lambda(k_\perp^\prime r^{}_\perp) \e^{\ii q_z z} \e^{\ii \varphi (\mu-\lambda)} | i \rangle.
 \end{aligned}
\end{equation}
Assuming the initial and final states are eigenstates of the projected OAM operator $L_z$ (\textit{i.e.} they have the form of Eq.~\eqref{eq:atomic_azimuthal}), the azimuthal factor can be integrated out:
\begin{align}
 \widetilde{\mathcal{M}}_{fi} &= \sum_{\lambda=-\infty}^{+\infty}
 \begin{aligned}[t]
  &\left(-\ii \e^{\ii \phi^\prime}\right)^\lambda \langle \beta | J_\mu(k^{}_\perp r^{}_\perp) J_\lambda(k_\perp^\prime r^{}_\perp) \e^{\ii q_z z} | \alpha \rangle \\
  &\times \int \diff{\varphi} \frac{\e^{\ii \varphi (m-m^\prime)}}{2\pi} \e^{\ii \varphi (\mu-\lambda)}
 \end{aligned} \notag \\
 &= \left(-\ii \e^{\ii \phi^\prime}\right)^{\mu + \Delta m} \langle \beta | J_{\mu}(k^{}_\perp r^{}_\perp) J_{\mu+\Delta m}(k_\perp^\prime r^{}_\perp) \e^{\ii q_z z} | \alpha \rangle \notag \\
 &= (-\ii)^{\mu+\Delta m} \e^{\ii \phi^\prime (\mu + \Delta m)} \mathcal{M}_{\beta \alpha}(\bs k^\prime, \bs k, \mu, \Delta m).
\end{align}
The change in the atomic electron's (projected) OAM, $\Delta m = m - m^\prime$, is the OAM transferred by the scattering electron in the collision.
The reduced matrix element is equal to:
\begin{equation} \label{eq:reduced_matrix_element}
 \mathcal{M}_{\beta \alpha} = \langle \beta | J_{\mu}(k^{}_\perp r^{}_\perp) J_{\mu+\Delta m}(k_\perp^\prime r^{}_\perp) \e^{\ii q_z z} | \alpha \rangle.
\end{equation}
This is now the only unknown.
Note that due to the form of $f_\mathrm{el}^V$ (see App.~\ref{app:elastic_scattering_amplitude}) and $\widetilde{\mathcal{M}}_{fi}$ (see above), the scattering amplitude's (\textit{i.e.} the outgoing wave's) azimuthal dependence is exactly $\e^{\ii (\ell+\Delta m) \phi^\prime}$, which implies the transfer of the scattering electron's OAM to the atomic state.

For the hydrogen wave functions (oriented with respect to the beam axis, and not the $\theta$-dependent momentum transfer $\bs q$ as in Sec.~\ref{sec:final_state}), Eq.~\eqref{eq:reduced_matrix_element} becomes a triple Bessel integral which can be formally solved using a collection of tricks.
The result is unfortunately unwieldy, and gives no further insight into the physics of the problem.

The next section provides a alternative analytical treatment that can be used to calculate the scattering amplitude for any specific transition, both numerically, and analytically.
This result is still useful though when analysing central scattering, for \textit{i.e.} $\theta=0$, Eq.~\eqref{eq:reduced_matrix_element} simplifies significantly, and due to the generality of the obtained expression, an OAM reciprocity theorem can be deduced.
These cases are discussed in Sec.~\ref{sec:central_scattering_amplitudes} and~\ref{sec:reciprocity}.

\subsection{Fourier representation} \label{sec:fourier}

The Fourier representation of a Bessel beam given by Eq.~\eqref{eq:bessel_fourier_representation} provides a solution for the problem of the relative coordinate described in Sec.~\ref{sec:general_formulation}.
Starting from Eq.~\eqref{eq:inelastic_scattering_amplitude_full} (with $N=(2\pi)^{5/2}$ as in ref.~\cite{VanBoxem_Rutherford}), and using Eqs.~\eqref{eq:inelastic_coulomb_potential} and~\eqref{eq:bessel_fourier_representation}, one arrives at the integral in question ($\ell^\prime$ is the outgoing beam's OAM, obscured by $f_{fi}^{PW}$ and the integral over the azimuthal Fourier coordinate):
\begin{equation} \label{eq:inelastic_amplitude_fourier}
 \begin{aligned}
  f_{fi}^\mathrm{V}(\bs k^\prime, \ell^\prime; \bs k, \ell) &= \frac{(-\ii)^\ell}{2\pi} \int \diff{\phi} \e^{\ii \ell \phi} f_{fi}^\mathrm{PW} (\bs q), \\
  &=-\frac{m_e e^2}{\pi\hbar^2} (-\ii)^\ell \int \diff{\phi} \frac{\e^{\ii \ell \phi}}{q^2} \langle f | \e^{\ii \bs q \cdot \bs r} - Z | i \rangle.
 \end{aligned}
\end{equation}
where the plane wave scattering amplitude, $ f_{fi}^\mathrm{PW}(\bs q)$ is given by Eq.~\eqref{eq:plane_wave_scattering_amplitude}.
This can be analytically calculated using the contour integration technique first described in ref.~\cite{VanBoxem_Rutherford}.
Several descriptive examples are treated explicitly below.

\begin{figure}
  \includegraphics{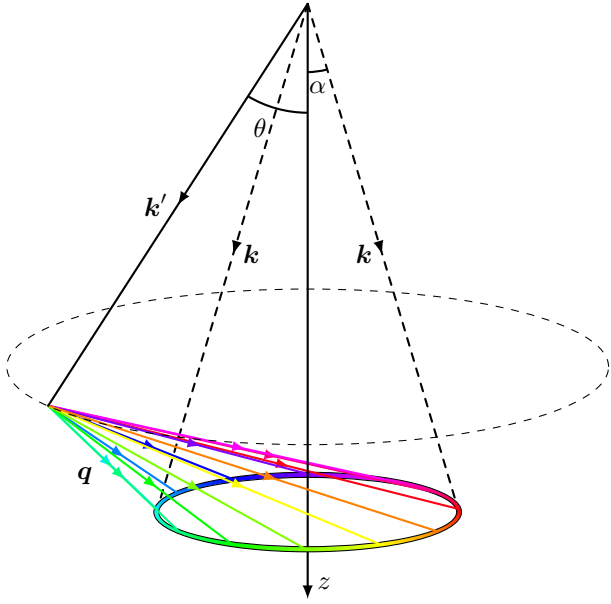}
  \caption{(color online) For cylindrical waves, and by extension vortex waves, there is no single momentum transfer. Instead, the various plane wave components of the cylindrical wave contribute to the final outgoing wave vector component. This is made explicit by the integral over the various momentum component vectors in Eq.~\eqref{eq:inelastic_amplitude_fourier}. A Bessel beam of $\ell=1$ is shown, and the relative phase of the various contributions is shown by the hue of the arrows. \label{fig:vortex_momentum_transfer}}
\end{figure}
\begin{figure}
  \includegraphics{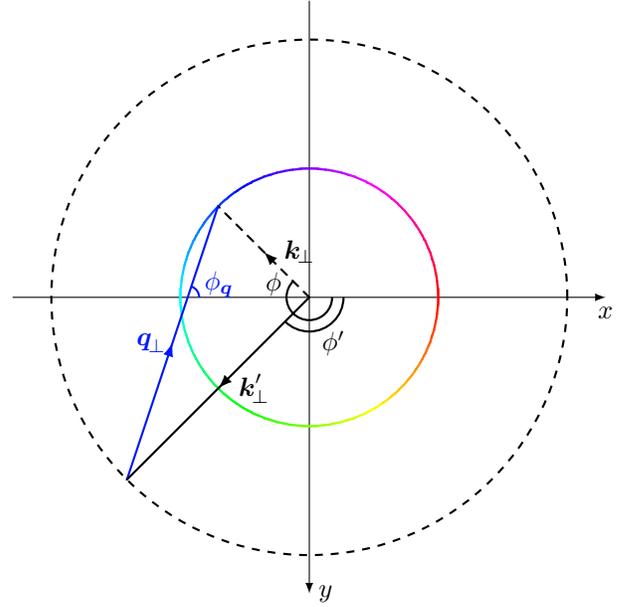}
  \caption{(color online) Top-down view of Fig.~\ref{fig:vortex_momentum_transfer}, showing the transverse plane scattering kinematics for an $\ell=1$ Bessel beam. \label{fig:vortex_transverse_momentum_transfer}}
\end{figure}

Eq.~\eqref{eq:inelastic_amplitude_fourier} has a direct physical interpretation, which is illustrated in Fig.~\ref{fig:vortex_momentum_transfer}.
For a certain scattering angle $\theta$ (with respect to the beam's principal direction) and energy $E^\prime = \hbar^2 k^{\prime 2}/(2m_e)$, each component plane wave of the incoming Bessel beam will \emph{coherently} interfere and contribute to the outgoing wave.
This makes it impossible to define a single momentum transfer $\bs q$ and thus also a unique final state orientation as discussed in Sec.~\ref{sec:final_state}.
So, in general, a $\phi$-dependent projection is required if one wants to discriminate final states defined with respect to the beam's axis.

To calculate the Bessel beam scattering amplitude analytically from Eq.~\eqref{eq:inelastic_amplitude_fourier}, one must first obtain Eq.~\eqref{eq:plane_wave_scattering_amplitude}, and then apply the following substitutions:
\begin{subequations} \label{eq:contour_substitutions}
 \begin{align}
  q^2 &= k_\perp^2 + k_\perp^{\prime 2} + q_z^2 - 2 k^{}_\perp k_\perp^\prime \cos{(\phi - \phi^\prime)}, \label{eq:contour_q} \\
  z &= \tan{\left(\frac{\phi - \phi^\prime}{2}\right)}, \\
  \cos(\phi - \phi^\prime) &= \frac{1-z^2}{1+z^2}, \\
  \e^{\ii \ell (\phi - \phi^\prime)} &= \left( \frac{\ii - z}{\ii + z} \right)^\ell, \\
  \diff{(\phi - \phi^\prime)} &= \frac{2\diff{z}}{1+z^2}.
 \end{align}
\end{subequations}
One can then extend the integration over $z$ (over the real axis) to an appropriate contour (such as an infinite semi-circle as previously used for elastic Coulomb scattering~\cite{VanBoxem_Rutherford}), and unleash the residue theorem.
As such, the $1s\rightarrow 1s$ (elastic) scattering amplitude is:
\begin{equation} \label{eq:scattering_amplitude_1s1s}
 \begin{aligned}
  f&_{1s \rightarrow 1s}^\mathrm{V} (\bs k^\prime, \ell; \bs k, \ell) \\
  &=
  \begin{aligned}[t]
   &-\frac{2 m_e e^2}{\hbar^2} (-\ii)^\ell \e^{\ii \ell \phi^\prime} \left( \frac{R_2 - R_1}{R_1 + R_2} \right)^{|\ell|} \\
   &\times \frac{a_0^2 R_1^2 R_2^2 + 2 Z^2 \left( R_1^2 + R_2^2 + 4 |\ell| R_1 R_2 \right)}{a_0^4 R_1^3 R_2^3}
  \end{aligned} \\
  & R_1^2 = \left(\frac{2 Z}{a_0} \right)^2 + q_z^2 + (k_\perp - k_\perp^\prime)^2 \\
  & R_2^2 = \left(\frac{2 Z}{a_0} \right)^2 + q_z^2 + (k_\perp + k_\perp^\prime)^2
 \end{aligned}
\end{equation}
Note that the outgoing OAM, $\ell^\prime$, is equal to the incoming OAM because none was transferred to the atomic electron.
Inelastic scattering amplitudes can also be calculated; below is the result for the $1s\rightarrow 2s$ transition:
\begin{equation} \label{eq:scattering_amplitude_1s2s}
 \begin{aligned}
  & f_{1s \rightarrow 2s}^\mathrm{V} (\bs k^\prime; \bs k, \ell) \\
  &=
  -\frac{\sqrt{2} m_e e^2}{\hbar^2} \left(\frac{Z}{a_0}\right)^4 (-\ii)^\ell \e^{\ii \ell \phi^\prime} \left( \frac{R_1 - R_2}{R_1 + R_2} \right)^{|\ell|} \\
  &\times \frac{3 (R_1^4 + R_2^4) + 6 |\ell| R_1 R_2 (R_1^2 + R_2^2) + 2 (1+ 2 \ell^2) R_1^2 R_2^2}{R_1^5 R_2^5} \\
  & R_1^2 = \left(\frac{3 Z}{2 a_0} \right)^2 + q_z^2 + (k_\perp - k_\perp^\prime)^2 \\
  & R_2^2 = \left(\frac{3 Z}{2 a_0} \right)^2 + q_z^2 + (k_\perp + k_\perp^\prime)^2
 \end{aligned}
\end{equation}

For non-spherical transitions (i.e. where $\Delta m \neq 0$), care must be taken about the final state orientation as discussed in Sec.~\ref{sec:final_state}.
Because there is now no unique momentum transfer $\bs q$, and there is both incoming and outgoing transverse momentum, the rotation over $\phi^\prime$ becomes one over $\phi_{\bs q}$ as shown in Figs.~\ref{fig:vortex_momentum_transfer} and~\ref{fig:vortex_transverse_momentum_transfer}.
As an example, we treat the $1s\rightarrow 2p$ transitions, for which the plane wave scattering amplitude is given by Eq.~\eqref{eq:plane_wave_scattering_amplitude_1s2p}.
For the rotation over $\phi_{\bs q}$, one can employ the complex representation of vector addition:
\begin{align}
 q_\perp \e^{\ii \phi_{\bs q}} &= k_\perp \e^{\ii \phi} - k_\perp^\prime \e^{\ii \phi^\prime} \notag \\
 &= \e^{\ii \phi^\prime}\left( k_\perp \e^{\ii (\phi-\phi^\prime)} - k_\perp^\prime \right).
\end{align}
This, together with Eq.~\eqref{eq:chi_in_function_of_q}, makes clear that the integrand in Eq.~\eqref{eq:inelastic_amplitude_fourier} is only dependent on $\phi-\phi^\prime$ and $q$ (which itself is only a function of that same variable, see Eq.~\eqref{eq:contour_q}).
For the $1s\rightarrow 2p_\pm$ transition, this gives:
\begin{equation} \label{eq:scattering_amplitude_1s2pplusmin}
 f_{1s\rightarrow 2p_\pm}^\mathrm{V} =
 \begin{aligned}[t]
  &-\frac{6\ii m_e e^2 (-1)^\ell}{\pi \hbar^2} \left( \frac{Z}{a_0} \right)^5 \e^{\ii (\ell \mp 1) \phi^\prime} \\
  &\times \int \diff{(\phi-\phi^\prime)} \e^{\ii \ell (\phi-\phi^\prime)} \frac{k_\perp \e^{\mp \ii (\phi-\phi^\prime)} - k_\perp^\prime}{q^2 \left[ q^2 + \left( \frac{3 Z}{2 a_0} \right)^2 \right]^3}.
 \end{aligned}
\end{equation}
Note that the outgoing wave has lost/gained OAM: the beam electron transfers OAM to the atomic state.
The integrated form is quite lengthy, so it is not shown here.
The remaining $2p_z$ state has the following scattering amplitude:
\begin{equation} \label{eq:scattering_amplitude_1s2pz}
 f_{1s\rightarrow 2p_z}^\mathrm{V} =
 \begin{aligned}[t]
  &-\frac{6\ii\sqrt{2} m_e e^2 (-1)^\ell}{\pi \hbar^2} \left( \frac{Z}{a_0} \right)^5 q_z \e^{\ii \ell \phi^\prime} \\
  &\times \int \diff{(\phi-\phi^\prime)} \e^{\ii \ell (\phi - \phi^\prime)} \frac{1}{q^2 \left[ q^2 + \left( \frac{3 Z}{2 a_0} \right)^2 \right]^3}.
 \end{aligned}
\end{equation}
The analytical expressions for these scattering amplitudes can be explicitly written down, but are too complex to present here and don't give much physical insight.

As an alternative to analytical calculation, Eq.~\eqref{eq:inelastic_amplitude_fourier} can be calculated \emph{numerically} as a function of scattering angle by using Eqs.~\eqref{eq:outgoing_momentum} and~\eqref{eq:chi_in_function_of_q}, and the following substitutions:
\begin{subequations} \label{eq:fourier_numerical_substitutions}
 \begin{align}
  q_\perp^2 &= k_\perp^2  + k^{\prime 2} \sin^2{\theta} - 2k_\perp k^\prime \sin{\theta} \cos{(\phi-\phi^\prime)}, \\
  q_z^2 &= (k_z - k^\prime \cos{\theta})^2, \label{eq:vortex_qz}
 \end{align}
\end{subequations}
and integrating over $\diff{(\phi - \phi^\prime)}$ numerically.
The functions in question are all in all well-behaved and standard numerical integration methods should have no issues with them.
The option of simple numerical integration is very useful if one wants to explore a different basis set of state wave functions, as long as they can be quantized in OAM as in Eq.~\eqref{eq:atomic_azimuthal}, and their scattering amplitude can be written as a function of $\phi-\phi^\prime$ explicitly.

Figures~\ref{fig:vortex_1s1s}, \ref{fig:vortex_1s2pz}, and \ref{fig:vortex_1s2pplusmin} show the scattering amplitudes for several characteristic transitions and different input beams.
In the limit $\ell=0$, $k_\perp=0$, which is the red, long-dashed line on each of the plots, the results of Fig.~\ref{fig:final_state}b are recovered.
The cross sections are discussed in more detail in Sec.~\ref{sec:hydrogen_cross_sections}

\section{Discussion} \label{sec:discussion}

\subsection{OAM Reciprocity} \label{sec:reciprocity}

\begin{figure}
 \centering
 \includegraphics{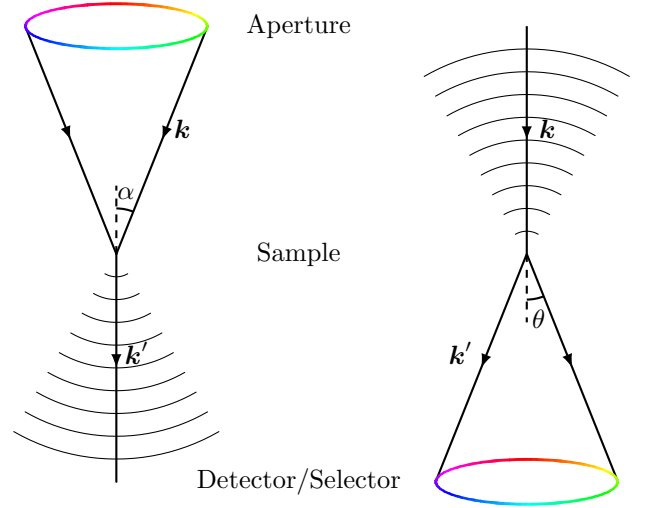}
 \caption{(color online) Representation of the symmetry exposed in Sec.~\ref{sec:reciprocity} between incoming and outgoing vortex waves in the case of incoming and outgoing plane waves, respectively. A specific transition has the same scattering amplitude at the scattering angle $\theta$ corresponding to the inverted geometry as shown here. \label{fig:inelastic_symmetry}}
\end{figure}

In the cylindrical scattering amplitude, Eq.~\eqref{eq:inelastic_scattering_amplitude_cylindrical}, one can consider two specific cases: incoming and outgoing non-vortex states.
For a fixed transition with a certain $\Delta m$, an incoming plane wave (so with $\ell=0$, $k_\perp=0$) will result in a scattered vortex wave with $\ell^\prime=\Delta m$.
Additionally, the summation in Eq.~\eqref{eq:inelastic_scattering_amplitude_cylindrical} can be performed using $J_\mu(k_\perp r_\perp) \xrightarrow{k_\perp\rightarrow 0} \delta_{\mu,0}$.
Using the explicit form of $f_\mathrm{el}^V$ (see App.~\ref{app:elastic_scattering_amplitude}), one obtains the relatively simple expression:
\begin{equation}
 f_{fi}^\mathrm{PW} (\bs k^\prime, \ell^\prime; k_z \bs e_z) =
 \begin{aligned}[t]
  &-\frac{2 m_e}{\hbar^2} (-\ii)^{\Delta m} \e^{-\ii \phi^\prime \Delta m} \frac{Ze^2}{k_\perp^2 + q_z^2} \\
  &\times \langle \beta | J_{\Delta m}(k_\perp^\prime r^{}_\perp) \e^{\ii q_z z} | \alpha \rangle.
 \end{aligned}
\end{equation}
Doing the same, but taking the outgoing wave to be a plane wave ($\ell^\prime=0$, $k_\perp^\prime=0$), one obtains a very similar expression:
\begin{equation} \label{eq:central_scattering_amplitude}
 f_{fi}^\mathrm{V} (\theta=0; \bs k, \ell) =
 \begin{aligned}[t]
  &-\frac{2 m_e}{\hbar^2} \frac{Ze^2}{k_\perp^2 + q_z^2} \delta_{\ell,\Delta m} \\
  &\times \langle \beta | J_{\Delta m}(k_\perp r_\perp) \e^{\ii q_z z} | \alpha \rangle.
 \end{aligned}
\end{equation}
What these two equations say is quite simple: an incident plane wave can gain/lose OAM by exciting a transition, and will then be scattered to a certain angle $\theta = \tan^{-1}(k_\perp^\prime/k_z^\prime)$.
However, a vortex wave with transverse $k_\perp$ will have the exact same probability of exciting that transition when scattering to $\theta=0$.
Note also that only when an incoming vortex beam's OAM matches the transition's $\Delta m$, can it be scattered to $\theta=0$ (see Sec.~\ref{sec:central_scattering_amplitudes}).
This can be seen as a form of \emph{reciprocity}~\cite{Potton,Deak,Findlay}, but in this case for inelastic scattering including OAM.

\subsection{Central scattering amplitudes} \label{sec:central_scattering_amplitudes}

Instead of considering the full $\theta$-dependent scattering amplitude, one can also only consider the $\theta=0$ case, for which the analytical expressions are much more tractable than for the complete calculation.
Eq.~\eqref{eq:central_scattering_amplitude} can be calculated directly using known integrals~\footnote{Found in Sec. 6.521 of ref.~\cite{GradshteynRyzhik}. Others can be derived by taking the derivative of both sides with respect to the parameters $a$ and $b$ in those equations.} of a set of Bessel $K$ and $J$ functions.
Alternatively, the analytical method from Sec.~\ref{sec:fourier} can be used to the same effect.

For $1s\rightarrow1s$, this gives:
\begin{equation} \label{eq:central_scattering_amplitude_1s1s}
 \begin{aligned}
  f&_{1s \rightarrow 1s}^\mathrm{V} (\theta=0;\bs k,\ell) \\
  &= \frac{2a_0 e^2 m_e \delta_{\ell,0}}{\frac{4Z^2}{a_0^2} + k_\perp^2 + q_z^2} \left(1 + \frac{\left(\frac{2Z}{a_0}\right)^2}{\left(\frac{2Z}{a_0}\right)^2 + k_\perp^2 + q_z^2}\right).
 \end{aligned}
\end{equation}
Note the extra term which comes from the first term in Eq.~\eqref{eq:inelastic_scattering_amplitude_full}.
If one compares this with the plane wave result in ref.~\cite{Schiff}, one immediately sees the symmetry of this expression if one replaces $k_\perp$ with $k_\perp^\prime$ and realizes $q^2 = k_\perp^{\prime 2} + q_z^2$ in this situation.
This result can also be obtained by setting $k_\perp^\prime=0$ in Eq.~\eqref{eq:scattering_amplitude_1s1s}.

For the $1s\rightarrow 2s$ transition, the following result is obtained:
\begin{equation}
 \begin{aligned}
  f&_{1s \rightarrow 2s}^\mathrm{V}(\theta=0;\bs k, \ell) \\
  &= \frac{8\sqrt{2}m_e e^2}{\hbar^2} \left( \frac{Z}{a_0}\right)^4 \frac{\delta_{\ell,0}}{\left[ k_\perp^2 + q_z^2 +\left(\frac{3 Z}{2 a_0}\right)^2 \right]}.
 \end{aligned}
\end{equation}
The Kronecker delta ensures the outgoing wave does not gain or lose OAM with respect to the incoming one, as only the $\ell=0$ mode can be non-zero at $\theta=0$.

By only considering central scattering, one can selectively measure a specific transition by preselecting a proper incoming vortex state.
This can be shown by considering an incoming $\ell=\pm 1$ vortex beam and $1s\rightarrow 2p_\pm$ atomic transitions.
The relevant central scattering amplitude is given by:
\begin{equation} \label{eq:central_scattering_amplitude_1s2ppm}
 \begin{aligned}
  f&_{1s \rightarrow 2p_\pm}^V(\theta=0;\bs k, \ell) \\
  &= - \frac{12 \ii m_e e^2}{\hbar^2} \frac{\delta_{\ell,\pm 1}}{k_\perp^2 + q_z^2} \left( \frac{Z}{a_0} \right)^5 \frac{k_\perp}{\left[ k_\perp^2 + q_z^2 + \left(\frac{3Z}{2a_0}\right)^2 \right]^3}.
 \end{aligned}
\end{equation}
For comparison, the central scattering amplitude for the $1s\rightarrow 2p_z$ transition is given below:
\begin{equation}
 \begin{aligned}
  f&_{1s \rightarrow 2p_z}^V (\theta=0; \bs k, \ell) \\
  &= \frac{12\ii\sqrt{2} m_e e^2}{\hbar^2} \frac{\delta_{\ell,0}}{k_\perp^2 + q_z^2} \left( \frac{Z}{a_0} \right)^5 \frac{q_z}{\left[k_\perp^2 + q_z^2 + \left( \frac{3Z}{2a_0}\right)^2 \right]^3}.
 \end{aligned}
\end{equation}
Note how the roles of $k_\perp$ and $q_z$ are reversed with respect to Eq.~\eqref{eq:central_scattering_amplitude_1s2ppm} (typical behavior for these $p$-character final states), and that the central $p_z$ scattering cross section is twice as large for the same parameters.
Take special note of the strict selection rule for $\theta=0$ expressed by the kronecker $\delta$'s, showing that the on-axis intensity for these transitions will be non-zero only for the right incoming beam.
Due to the summation in the full expression, Eq.~\eqref{eq:inelastic_scattering_amplitude_cylindrical}, a mixture of outgoing vortex waves will generally be emitted from each scattering event regardless its $\Delta m$.
The rate at which these various components contribute is determined by the weighting expressed by that equation, which is not trivial.

Lastly, it is important to note that the three $2p$ states considered here are degenerate unless a magnetic field is applied, which adds a Zeemann energy, splitting the non-zero OAM levels with the magnetic field.
Larger fields will also induce spin-orbit coupling in an atomic system, further complicating the wave functions and interactions involved.~\cite{Galindo,Popov}

\subsection{Hydrogen scattering amplitudes} \label{sec:hydrogen_cross_sections}

\begin{figure}
 \centering
 \includegraphics[width=\linewidth]{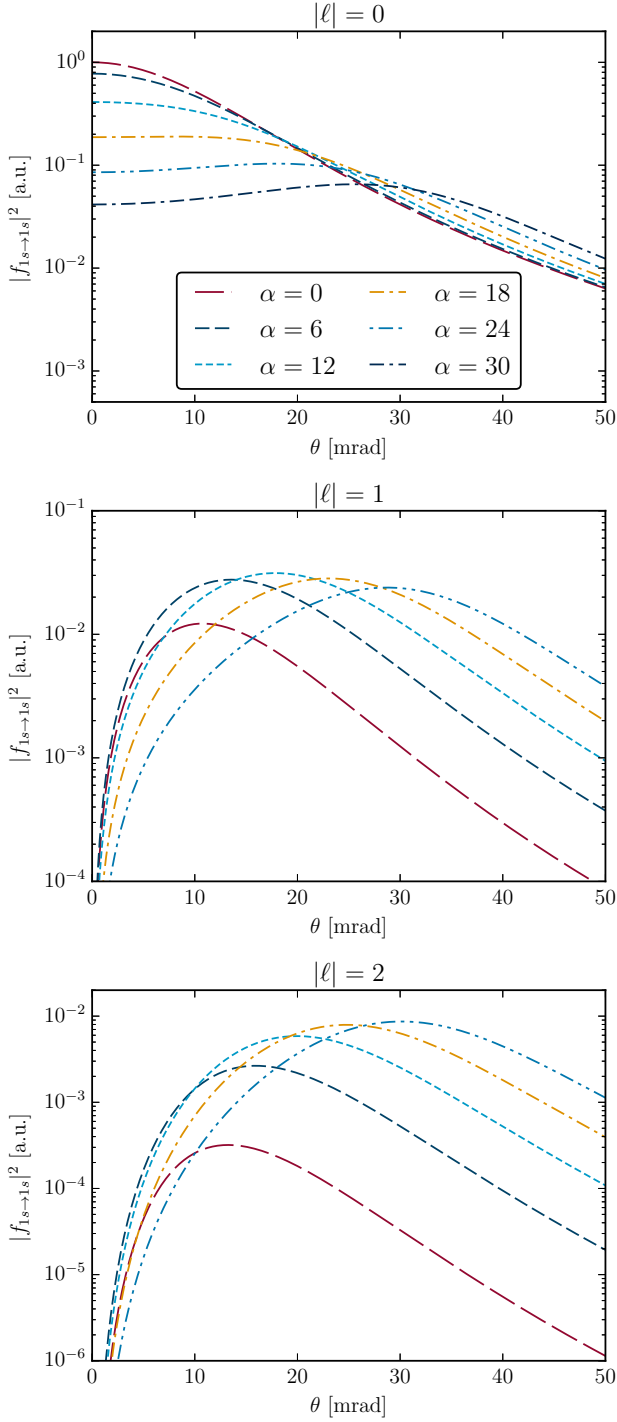}
 \caption{(color online) $1s\rightarrow 1s$ (elastic) hydrogen differential scattering cross sections for electron Bessel beams of different OAM and transverse momentum, denoted by the Bessel beam opening angle $\alpha$ (in mrad). Similar features as for the screened Coulomb scattering amplitude are present~\cite{VanBoxem_Rutherford}. \label{fig:vortex_1s1s}}
\end{figure}
\begin{figure}
 \centering
 \includegraphics[width=\linewidth]{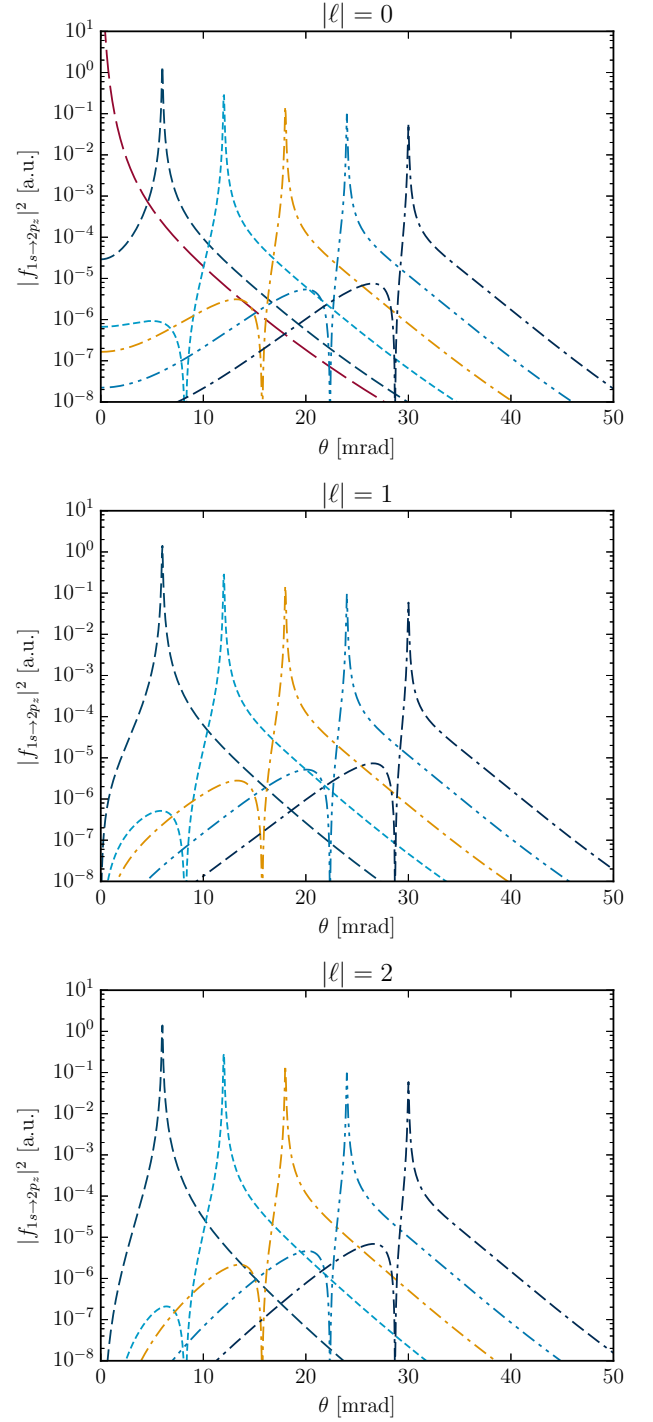} 
 \caption{(color online) The scattering cross sections for excitation of the $2p_z$ final state. Note that the peaked maximum is smooth when viewed on the much smaller $\theta$ scale of \textit{e.g.} Fig.~\ref{fig:final_state}b. The small maximum in the region where $\theta<\alpha$, although real, is relatively insignificant when compared to the cross section's maximum value. \label{fig:vortex_1s2pz}}
\end{figure}
\begin{figure*}
  \centering
  \includegraphics[width=\linewidth]{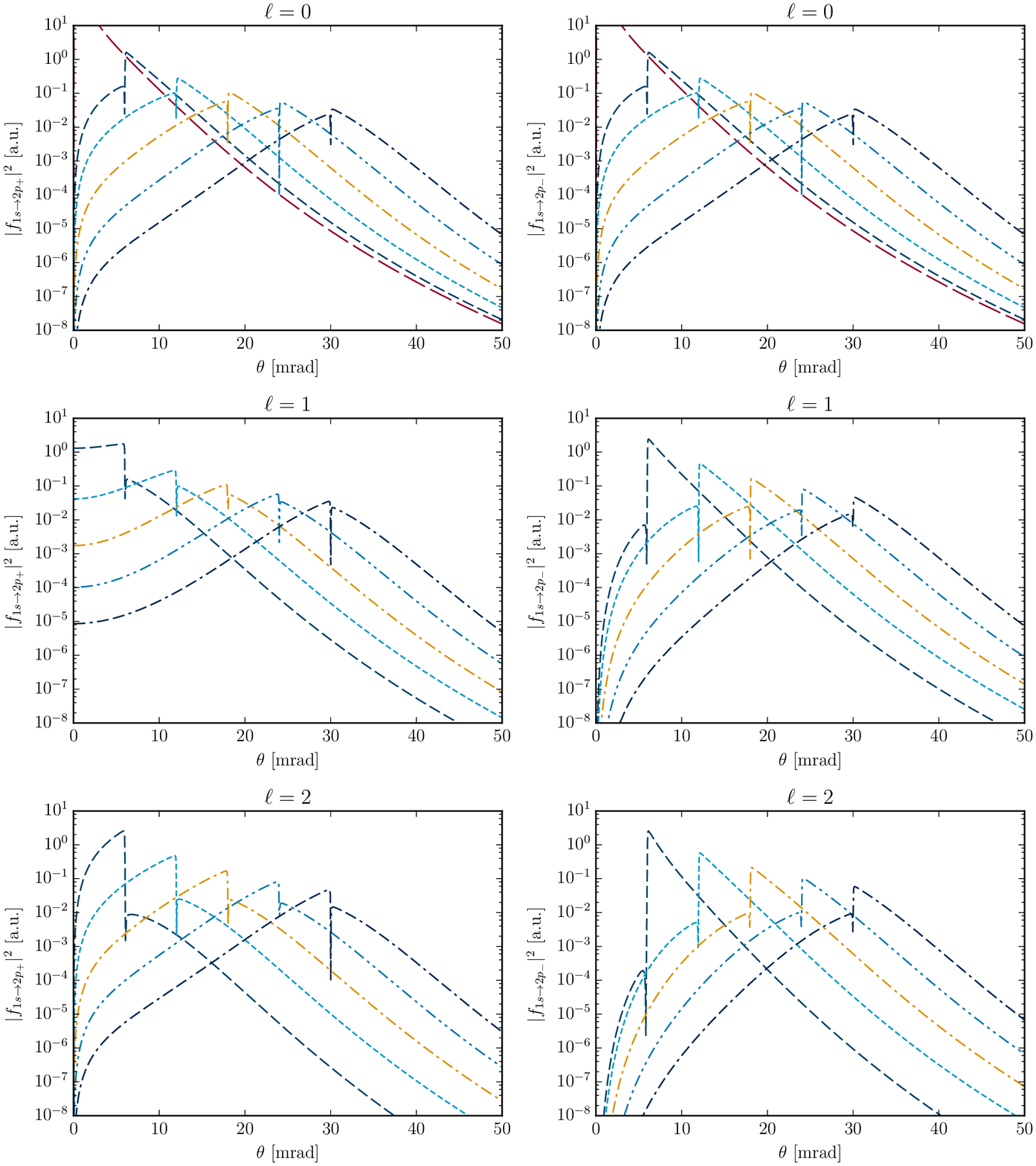}
  \caption{(color online) The scattering cross sections for OAM transfer events to the $2p_\pm$ states are shown for positive incoming beam OAM, $\ell$. The negative $\ell$ results are omitted because they are identical to these with $2p_+$ and $2p_-$ swapped due to the symmetry between both situations. Legend is identical to the one in Fig.~\ref{fig:vortex_1s1s}. \label{fig:vortex_1s2pplusmin}}
\end{figure*}

Several properties of the scattering amplitudes presented in \ref{sec:fourier}, Eqs.~\eqref{eq:scattering_amplitude_1s1s}-\eqref{eq:scattering_amplitude_1s2pz}, are immediately apparent by only looking at the equations themselves:
\begin{itemize}
 \item Higher energy levels (\textit{i.e.} larger $n$) introduce higher order $\ell$ dependence.
 \item Transitions between spherically symmetric states (for which the scattering amplitude is independent of $\phi$) only depend on the magnitude of $\ell$, and not its sign. This is expected as everything involved is symmetric with respect to rotations around the $z$ axis. Non-symmetrical transitions depend strongly on the value of $\ell$.
 \item The scattering electron wave transfers OAM to the atomic state, and this transfer is reflected in the outgoing wave.
 \item Taking the plane-wave limit ($\ell=0$, $k_\perp\rightarrow 0$), one recovers the usual plane wave scattering amplitudes (see \textit{e.g.} Ref.~\cite{Schiff} and Fig.~\ref{fig:final_state}b).
\end{itemize}

The results of Sec.~\ref{sec:fourier} are also shown in Figs.~\ref{fig:vortex_1s1s}-\ref{fig:vortex_1s2pplusmin} (in atomic units).
The analytical results in Eqs.~\eqref{eq:scattering_amplitude_1s1s}, \eqref{eq:scattering_amplitude_1s2s}, \eqref{eq:scattering_amplitude_1s2pz}, and \eqref{eq:scattering_amplitude_1s2pplusmin}, and the numerically integrated scattering amplitudes using Eq.~\eqref{eq:fourier_numerical_substitutions} were confirmed to be equal.
These results and their physical implications shall now be discussed.
Remember that the differential cross section is the probability of an electron being detected at a certain scattering angle, which is the primary interpretation that will be given here.
Note that the plots show only a line profile; the scattering amplitudes shown here are all cylindrically symmetrical (up to a possible vortex phase factor), resulting in the scattered electrons forming doughnut-shaped intensity profiles.

The elastic differential scattering cross section, shown in Fig.~\ref{fig:vortex_1s1s}, is very similar to that of the screened Coulomb potential previously treated~\cite{VanBoxem_Rutherford}.
For a non-vortex beam, $\ell=0$, higher transverse momentum shifts this peak off-center, although even for an incoming hollow beam (with no low-$k_\perp$ components, generated by \textit{e.g.} an annular aperture), a significant amount of intensity is still scattered on-axis.
This can be deduced from the additive contributions of various transverse momentum components to the scattering amplitude (shown in the topmost plot of Fig.~\ref{fig:vortex_1s1s}), which additively result in a non-zero on-axis differential scattering cross section.
For beams containing a phase vortex, \textit{i.e.} $\ell\neq 0$, there is no on-axis intensity and the angle at which the cross section peaks shifts outwards with increasing incoming transverse momentum.
This feature could be exploited as a sensitive detector for non OAM preserving transitions in future experiments.
Another spherically symmetric transition is $1s\rightarrow 2s$.
Its inelastic differential scattering cross section (given explicitly by Eq.~\eqref{eq:scattering_amplitude_1s2s}) has the same shape and behavior as the $1s\rightarrow 1s$ cross section, and is therefore not explicitly shown.

The differential cross sections for ground state excitations to various $2p$ substates (quantized along the beam direction) are shown in Figs.~\ref{fig:vortex_1s2pz} and~\ref{fig:vortex_1s2pplusmin}.
The $2p_z$ cross section shows a narrow peak that moves away from the beam axis, giving way for a zero that appears at the threshold transverse momentum defined by  the condition $q_z=0$ (which is a prefactor in Eq.~\eqref{eq:scattering_amplitude_1s2pz} and is located at:
\begin{equation}
 \theta_0 = \cos^{-1}{\left(\frac{k}{k^\prime} \cos{\alpha} \right)} < \alpha.
\end{equation}

When an incoming vortex beam transfers OAM to an atomic electron, the associated differential scattering amplitude reflects this, as shown in Fig.~\ref{fig:vortex_1s2pplusmin}.
In general, the lower the outgoing OAM is in magnitude $|\ell^\prime|$, the higher the scattering cross section is for low angles.
Another general feature of these results is that transitions where OAM is given to the atomic electron, $|\ell^\prime|<|\ell|$, have a relatively larger differential cross section for scattering angles $\theta < \alpha$.
Note that not all OAM needs to be transfered for this to be visible, and these differences are orders of magnitude larger for higher outgoing OAM.
Transitions where OAM is taken from the atomic electron, $|\ell^\prime|>|\ell|$, have larger differential cross sections for larger angles, $\theta > \alpha$.
This can be clearly seen by comparing  the plots in the left column of Fig~\ref{fig:vortex_1s2pplusmin} to those in the right one (the top row only shows the latter case, $|\ell^\prime| > |\ell|$).
Perhaps the most evident form of this is that an $\ell=1$ beam exciting a $1s\rightarrow 2p_+$ transition will scatter most electrons on the beam axis and even to $\theta=0$.
This is possible because for this specific transition, the outgoing beam has lost its OAM, and does not suffer from the phase singularity previously forcing its amplitude to zero there.
The same happens for an $\ell=-1$ beam exciting a $1s\rightarrow 2p_-$ transition, as the scattering amplitudes are identical.
For higher order incoming beams, this change is less dramatic, as the outgoing vortex phase still forces the central scattering to zero.
Although it is not displayed clearly on these figures' scales, the (red, long-dashed) plane wave differential cross section has the same shape as in Fig.~\ref{fig:final_state}b, and its central zero gets pushed outwards for higher transverse momentum.
The sharp dip at around $\alpha=\theta$ is exactly the zero of $f_{1s\rightarrow 2p_\pm}^{PW}$ in Fig.~\ref{fig:final_state}b at $\theta=0$, but shifted outwards due to the non-zero incoming transverse momentum.
It coincides with the peak for the $2p_z$ differential cross section at the same scattering angle.
These shifts of intensity can be understood intuitively by considering the transverse profiles of a vortex beam, where higher OAM is generally paired with a larger spatial extent of the wave function.
Finally, the large angle scattering of both situations also shows a quantitative offset which could also be used to differentiate the events.

\section{Conclusions}

We extended inelastic quantum scattering theory to non-trivial incoming electron waves, including orbital angular and transverse momentum.
A quick review of the textbook theory illuminated some often forgotten, but important facts about the final state and momentum transfer.
Two different methods were then applied to obtain the vortex scattering amplitudes of inelastic transitions for a hydrogen-like system.
Special attention was given to the atomic state's OAM and the consequences of it being changed by a scattering electron.

The first method involves the Bessel addition theorem, which leads to unwieldy analytical expressions for all scattering amplitudes.
Nonetheless, these calculations led to simple selection rules when OAM transfer is involved and allows to estimate the regime of scattering angles for which they are valid.
Additionally, a form of OAM reciprocity was shown to exist, tied to the OAM transfer and the central scattering amplitude.

The second approach, using the Fourier representation of the Bessel beam, resulted in a purely analytical method to obtain hydrogen scattering amplitudes, along with a less cumbersome numerical solution using an intermediate result.
The scattering amplitudes are influenced strongly in the presence of OAM transfer, even outside of the ideal selection rule validity regime ($\theta=0$).
Combined with energy filtering, the predicted asymmetry could provide a means to better separate scattering contributions for various final states with distinct OAM, leading to an improved measurement of the final state density.
This, in turn, would provide atomic-resolution magnetic information.
Specific $\Delta m$ transitions can be filtered from the total scattered intensity using any combination of pre- or post-selection of OAM of the scattering electron, and incoming and outgoing transverse momentum (\textit{i.e.} by limiting collection and convergence angles).
The selection on transverse momentum is routinely done by choosing detector geometry, selected area apertures, and objective apertures.
Post-selection on OAM has yet to be practically implemented efficiently in electron microscopes.
Once this is in place, though, the use of OAM selectors will improve selectivity to certain transitions significantly, as a large amount of background signal caused by electrons with the ``wrong" OAM will be eradicated.

\acknowledgments
RVB acknowledges support of the FWO (Fonds Wetenschappelijk Onderzoek - Vlaanderen) as Aspirant.
JV acknowledges financial support from the EU under the Seventh Framework Program (FP7) under a contract for an Integrated Infrastructure Initiative, Reference No. 312483-ESTEEM2, and the European Research Council under the FP7, ERC Starting Grant 278510 VORTEX.
Figures were made with TikZ, NumPy and Matplotlib~\cite{TikZ,NumPy,Matplotlib}.

\appendix

\section{Elastic vortex coulomb scattering amplitude} \label{app:elastic_scattering_amplitude}

In the cylindrical expansion of the inelastic scattering amplitudes, the elastic scattering amplitude appears.
This was calculated in ref.~\cite{VanBoxem_Rutherford}, and the result is repeated here:
\begin{equation} \label{eq:elastic_scattering_amplitude}
 \begin{aligned}
  f_\mathrm{el}^V(\bs k^\prime; \bs k, \ell) =&~ -\frac{2m_e V_0}{\hbar^2} \frac{\ii^\ell \e^{\ii \ell \phi^\prime}}{r_1 r_2} \left( \frac{r_1-r_2}{r_1+r_2} \right)^{|\ell|} , \\
  r_1^2 =&~ q_z^2 + \mu^2 + (k^{}_\perp - k_\perp^\prime)^2, \notag \\
  r_2^2 =&~ q_z^2 + \mu^2 + (k^{}_\perp + k_\perp^\prime)^2. \notag
 \end{aligned}
\end{equation}
Here, $V_0$ is either $Z e^2$ or $e^2$ depending on the context, and the treatment in this paper deals with unscreened Coulomb potentials only, so $\mu=0$ everywhere.

\bibliography{inelastic_electron_vortex_scattering}

\end{document}